\pdfmapline{-dummy CMR12}

\documentclass[
 reprint,
superscriptaddress,
 amsmath,amssymb,
 aps,
prb,
citeautoscript,
]{revtex4-2}

\usepackage{physics}
\usepackage{siunitx}
\usepackage[font=small,skip=3pt, belowskip=-10pt]{caption}
\usepackage{multirow}
\usepackage{hyperref}
\usepackage{float}
\usepackage{gensymb}

\usepackage{parskip}
\usepackage{graphicx}
\usepackage{csquotes}
\def\bigO#1{\mathcal{O}(#1)}

\usepackage{xcolor}

\usepackage{makecell}

\newcommand{\e}{\textrm{e}}

\begin{document}
\preprint{APS/123-QED}

\title{Towards Efficient Quantum Computing for Quantum Chemistry: 
	Reducing Circuit Complexity with Transcorrelated and Adaptive Ansatz Techniques}

\author{Erika Magnusson}
	\affiliation{
	Department of Chemistry and Chemical Engineering, 
	Chalmers University of Technology, 41296 Gothenburg, Sweden
}

\author{Aaron Fitzpatrick}
\affiliation{Algorithmiq Ltd, Kanavakatu 3C, FI-00160 Helsinki, Finland}

\author{Stefan Knecht}
\affiliation{Algorithmiq Ltd, Kanavakatu 3C, FI-00160 Helsinki, Finland}

\author{Martin Rahm}
 \email{martin.rahm@chalmers.se}
	\affiliation{
	Department of Chemistry and Chemical Engineering, 
	Chalmers University of Technology, 41296 Gothenburg, Sweden
}

\author{Werner Dobrautz}
 \email{werner.dobrautz@gmail.com}
 	\affiliation{
 	Department of Chemistry and Chemical Engineering, 
 	Chalmers University of Technology, 41296 Gothenburg, Sweden
 }

\date{\today}

\begin{abstract}

The near-term utility of quantum computers is hindered by hardware constraints in the form of noise. One path to achieving noise resilience in hybrid quantum algorithms is to decrease the required circuit depth -- the number of applied gates -- to solve a given problem.
This work demonstrates how to reduce circuit depth by combining the transcorrelated (TC) approach with adaptive quantum ansätze and their implementations in the context of variational quantum imaginary time evolution (AVQITE).
The combined TC-AVQITE method is used to calculate ground state energies across the potential energy surfaces of H$_4$, LiH, and H$_2$O. In particular, H$_4$ is a notoriously difficult case where unitary coupled cluster theory, including singles and doubles excitations, fails to provide accurate results. Adding TC yields energies close to the complete basis set (CBS) limit while reducing the number of necessary operators -- and thus circuit depth -- in the adaptive ansätze. The reduced circuit depth furthermore makes our algorithm more noise-resilient and accelerates convergence. Our study demonstrates that combining the TC method with adaptive ansätze yields compact, noise-resilient, and easy-to-optimize quantum circuits that yield accurate quantum chemistry results close to the CBS limit.

\end{abstract}


\maketitle

\section{Introduction}

The challenge at the heart of quantum chemistry is the electronic structure problem. 
This problem, encapsulated in the Schrödinger equation, scales exponentially with system size. 
Numerous computational approaches exist for tackling this challenge, ranging from approximate mean-field theories like Hartree-Fock (HF),~\cite{Helgaker_2000} more accurate but 
costly methods like coupled cluster (CC)~\cite{Cicek1966, Bartlett2007}, density matrix renormalization group (DMRG)~\cite{White1992, Chan2011, Baiardi2020a} and quantum Monte Carlo (QMC) methods~\cite{Nightingale1998, Becca2017, Guther2020}, 
to exact, but exponentially-scaling, full configuration interaction (FCI)/exact diagonalization (ED). 
In recent years, attempts have been made to circumvent the unfavourable scaling of highly accurate quantum chemistry using quantum computers. 
Quantum hardware is believed to be particularly well suited for simulating quantum systems like molecules and may enable a significant computational speedup \cite{bauer_quantum_2020, mcardle_quantum_2020}.
However, given the existence of conventional numerical methods that have been refined over decades, it is still uncertain if quantum algorithms can provide a genuine quantum advantage over established techniques.\cite{Liu_2022, lee_evaluating_2023, gonthier_measurements_2022, Bittel2021}

Unfortunately, noise severely limits practicable circuit depths on current and near-term quantum processors.
Furthermore, the number of qubits needed to encode quantum chemistry on quantum hardware is proportional to the basis set size or the number of orbitals in the case of an active space approach. 
Thus, the achievable accuracy on quantum hardware is severely limited as either small, often minimal, basis sets have to be used or calculations must be done with very small active spaces to fit the problem on current quantum hardware.
Despite these constraints, quantum hardware may, in the future, outperform conventional computation in specialized instances, such as modelling highly correlated systems \cite{cao_quantum_2019}.

Various algorithms have been devised to advance toward practical quantum advantage in the current noisy intermediate-scale quantum (NISQ) regime. Most of these NISQ algorithms are variational, i.e., based on the variational theorem. Variational quantum algorithms (VQAs)~\cite{Cerezo2021, Yuan2019} can significantly reduce quantum circuit depth by offloading calculations that do not strictly need quantum properties to a conventional computer. This idea follows naturally from trying to use the quantum computer as little as possible. 
VQAs are heuristic and rely on an ansatz circuit, which is optimized following some scheme. A considerable drawback of VQAs is that many measurements are needed for this optimization procedure, a factor that may limit or remove the chances for practical quantum advantage.~\cite{gonthier_measurements_2022} Despite this drawback, for reasons related to the limitations of current hardware, VQAs are by far the most investigated type of quantum algorithm to date.
The variational quantum eigensolver (VQE) \cite{peruzzo_variational_2014, Tilly2022} is the most well-known VQA. However, other methods, such as variational quantum imaginary time evolution (VarQITE), are competitive alternatives \cite{mcardle_variational_2019}.

A myriad variations of and additions to these VQAs have been made to improve them in search of practical quantum advantage. A non-exhaustive list of such approaches includes reducing circuit depth by gradually building the ansatz circuit to be only as deep as needed~\cite{grimsley_adaptive_2019, tang_qubit-adapt-vqe_2021, gomes_adaptive_2021, Grimsley2023, Feniou2023, Chen_Gomes_Niu_de_Jong_2024}, reducing qubit requirements by similarity transforms~\cite{Motta2020, schleich2021improving,  Kumar2022, sokolov_orders_2022, dobrautz_ab_2023, Huang2023a, Bauman2019, Bauman2019b, NicholasP2021, Bauman2022}, or post-processing \cite{bierman_improving_2023, mejuto-zaera_quantum_2023}.
Among these additions, explicitly correlated methods~\cite{Hylleraas1929, Httig2011, Kong2011, Tenno2011, Tenno2012, Grneis2017, Kutzelnigg1985, Tenno2004, Tenno2004b} like the transcorrelated (TC) method~\cite{Boys1969, Handy1969, Handy1969b, dobrautz_compact_2019, Cohen2019, Guther2021, Baiardi2020, Baiardi2022, Liao2023, Liao2021, Schraivogel2023, Schraivogel2021, Ammar2022} make it possible to obtain more accurate results with smaller basis sets by incorporating the problematic electronic cusp condition~\cite{Kato1957} into the Hamiltonian. The TC approach also has the added benefit of providing a more compact ground state wavefunctions.~\cite{dobrautz_compact_2019}
A consequence of this compactness is that the ground state of the TC Hamiltonian is easier to prepare with shallower quantum circuits~\cite{Mcardle2020, sokolov_orders_2022, dobrautz_ab_2023}.

Explicitly correlated and TC-based approaches have also recently been applied to increase the accuracy while lowering the resource requirements of quantum chemistry calculations on quantum hardware.~\cite{Motta2020, schleich2021improving, Mcardle2020, sokolov_orders_2022, dobrautz_ab_2023, Kumar2022, Volkmann2024}
	Motta \textit{et al.} employed canonical transcorrelated F12 (CT-F12) theory\cite{Motta2020} to accurately calculate the ground state energy of various small molecular systems with fewer required qubits. 
	Kumar \textit{et al.} extended CT-F12 theory on quantum hardware to excited state energies,\cite{Kumar2022}
	and Schleich and coworkers used an explicitly correlated a posteriori correction\cite{schleich2021improving} to improve ground state energy estimates from VQE calculations. 
	McArdle and Tew\cite{Mcardle2020} have merged the TC and VarQITE methods to study small Hubbard lattices, utilizing the improved compactness of the ground state solution due to the TC approach.\cite{dobrautz_compact_2019}
	Some of the present authors have developed an optimized TC-VarQITE approach\cite{sokolov_orders_2022} for \textit{ab initio} problems that drastically reduces the necessary qubit number to obtain accurate spectroscopic data of small molecular systems on quantum hardware.\cite{dobrautz_ab_2023}
	Our previous work relied on pre-determined and fixed quantum circuit ansätze like the unitary coupled cluster (UCC)\cite{Sokolov2020, Anand2022} or hardware efficient ansätze\cite{kandala_hardware-efficient_2017}. 
	However, full UCC ansätze are not a viable option for current noisy quantum hardware due to their required long circuit depths.
	On the other hand, hardware efficient ansätze can have convergence problems due to their heuristic nature and lack of chemical/physical motivation.

In this work, we present an extension of the TC-VarQITE approach by combining Gomes \textit{et al.}'s adaptive variational quantum imaginary time evolution algorithm (AVQITE)~\cite{gomes_adaptive_2021} with the TC method.
	The rationale for this approach is that the increased compactness of the ground state wavefunction due to the TC method\cite{dobrautz_compact_2019, sokolov_orders_2022} should lead to shallower adaptive quantum circuits.
The capability and strength of the resulting algorithm, transcorrelated adaptive variational quantum imaginary time evolution (TC-AVQITE), is then evaluated through simulations of near-term quantum devices.

This paper is structured as follows. First, we discuss in Section \ref{section:theory} the constituent parts of the TC-AVQITE algorithm and introduce relevant terminology. Next, we detail the implementation of TC-AVQITE in Section \ref{sec:method}, and provide computational details for the numerical studies. After discussing the numerical data in Section \ref{section:results}, we conclude, address possible improvements, and outline future work.

\section{Theory}\label{section:theory}
TC-AVQITE is built upon multiple methods and algorithms. To begin with, AVQITE is a combination of adaptive ansätze \cite{grimsley_adaptive_2019, tang_qubit-adapt-vqe_2021} and VarQITE \cite{mcardle_variational_2019, Yuan2019}. VarQITE is, in turn, a variational rephrasing of quantum imaginary time evolution (QITE) \cite{motta_determining_2019, Huang2023, Nishi2021, Tsuchimochi2023, Cao2022Quantumimaginarytime}. In what follows, we briefly introduce the electronic structure problem, followed by QITE, VarQITE, adaptive ansätze, and AVQITE. To conclude, we describe the TC method.

\subsection{The Electronic Structure Problem}

The electronic structure problem can often be reduced to solving the non-relativistic Schrödinger equation, 
either in stationary form, 
\begin{equation}\label{eq:schrodinger-stat}
	\Hat{H}\ket{\psi} = E\ket{\psi},
\end{equation}
with the system's Hamiltonian $\hat H$, eigenstates $\ket{\psi}$, and corresponding eigenenergies $E$;
or in time-dependent form (in other words, a dynamics simulation)
\begin{equation}\label{eq:schrodinger-time}
	\hat H \ket{\psi(t)} = i \frac{d}{dt}\ket{\psi(t)}.
\end{equation}
% $\Hat{H}\ket{\psi(\Vec{r}, t)} = E\ket{\psi(\Vec{r}, t)}$. 
%$\Hat{H}\ket{\psi} = E\ket{\psi}$. 
Decoupling electronic and nuclear degrees of freedom is often justified by invoking the Born-Oppenheimer approximation. 
When expressed within second quantization, the electronic Hamiltonian then reads as
\begin{equation}
	\hat{H} = \underbrace{\sum_{pq} h_{p}^q a^\dagger_p a_q}_\text{one-body terms} + \underbrace{\frac{1}{2} \sum_{pqrs} V_{pq}^{rs} a^\dagger_p a^\dagger_q a_r a_s}_\text{two-body terms},
	\label{eq:2bodyterms}
\end{equation}
where $a^{(\dagger)}_i$ is the annihilation (creation) operator of an electron in spin-orbital $i$, with the integrals
\begin{equation}
	h_{p}^q = \int \phi_p^*(\Vec{x}) \left( -\frac{\nabla^2}{2}- \sum_{i,I} \frac{Z_I}{|\Vec{r_i}-\Vec{R_I}|} \right) \phi_q(\Vec{x}) \dd \Vec{x}
	\label{eq:onebody}
\end{equation}
and
\begin{equation}
	V_{pq}^{rs} = \int \frac{\phi_p^*(\Vec{x}_1) \phi^*_q(\Vec{x}_2)\phi_r(\Vec{x}_1)\phi_s(\Vec{x}_2)}{|\Vec{r}_i - \Vec{r}_j|} \dd \Vec{x}_1 \dd \Vec{x}_2,
	\label{eq:twobody}
\end{equation}
where $\phi(\Vec{x})$ are the basis functions, $Z_I$ the charge number, $\Vec{r}_i$ the electron positions, $\Vec{R}_I$ the nucleon positions. From the shape of these integrals, we can note that the \textquote{one-body terms} include the kinetic energy and nuclear repulsion, while the \textquote{two-body terms} describe the electron-electron interaction.

\subsection{Imaginary time evolution, QITE and VarQITE}
QITE~\cite{motta_determining_2019, Huang2023, Nishi2021, Tsuchimochi2023, Cao2022Quantumimaginarytime} is a quantum computer implementation of imaginary time evolution (ITE),~\cite{vonderLinden1992, Ceperley1995, Trivedi1990, Guther2020} a method used in various fields of science, such as statistical mechanics, cosmology, and quantum mechanics.\cite{mcardle_variational_2019} 
ITE works by expressing the time-dependent Schrödinger equation, Eq.~\eqref{eq:schrodinger-time}, as dependent on imaginary time instead of time, $t \to i \tau$, in the so-called Wick-rotated form\cite{wick_properties_1954}
\begin{equation} \label{eq:wick}
	\pdv{}{\tau}\ket{\psi(\tau)} = - \Hat{H}\ket{\psi(\tau)}. %\\    
	%\begin{split}
	%    \pdv{}{\tau}\ket{\psi(\tau)} =& - (\Hat{H} -E_\tau)\ket{\psi(\tau)}, \\    
	%  \text{with }  E_\tau =& \bra{\psi(\tau)} \Hat{H} \ket{\psi(\tau)}.
	%\end{split}
\end{equation}
By integrating Eq.~\eqref{eq:wick} and given an initial state, $\ket{\psi(0)}$, one can obtain the state $\ket{\psi(\tau)}$ for any imaginary time $\tau$ as
\begin{equation} \label{eq:qite}
	\ket{\psi(\tau)} = \frac{\textrm{e}^{-\Hat{H}\tau} \ket{\psi(0)}}{\sqrt{\bra{\psi(0)} \textrm{e}^{-2\Hat{H}\tau} \ket{\psi(0)}}}.
\end{equation}
As $\tau \to \infty$, the state $\ket{\psi(\tau)}$ converges to the ground state of the Hamiltonian $\Hat{H}$, given that the initial state, $\ket{\psi(0)}$, overlaps with the ground state \cite{mcardle_variational_2019}. Fortunately, for quantum chemistry problems, this requirement is usually not particularly restrictive as easily preparable states with (in most cases) non-vanishing overlap to the ground state exist. 
One example is the wave function obtained by solving the Hartree-Fock equations. 
However, counterexamples to this assumption of easily preparable states with non-vanishing overlap exist\cite{lee_evaluating_2023}.
%, the state by physicists known as the Hartree-Fock state. 

ITE is a so-called projector method related to the power,\cite{Mises1929} Lanczos\cite{Lanczos1950} and Davidson method\cite{Davidson1975}, 
which yields the ground state of a system by repeated application of the operator $\textrm{e}^{-\hat H \tau}$ on the initial state $\ket{\psi(0)}$. 
Consequently, ITE does not rely on the variational principle and thus can be used to obtain ground states of non-Hermitian Hamiltonians, as present in open quantum systems\cite{Feshbach1958, Kamakari2022, Chen_Gomes_Niu_de_Jong_2024}, transport problems\cite{Beenakker1997, May2011}, and the transcorrelated method\cite{sokolov_orders_2022, Mcardle2020}. 

To perform ITE on a quantum computer (i.e., QITE), the exponential $\e^{-\Hat{H}\tau}$ is approximated by its Taylor series for a small imaginary time step $\Delta \tau$.~\cite{motta_determining_2019}
Implementing QITE on quantum hardware is not straightforward because the operator $\e^{-\Hat{H}\tau}$ is non-unitary. 
Consequently, $\e^{-\Hat{H}\tau}$ must be approximated by unitary operations, which can require deep quantum circuits\cite{motta_determining_2019}.

\begin{figure}
	\includegraphics[width=0.33\textwidth]{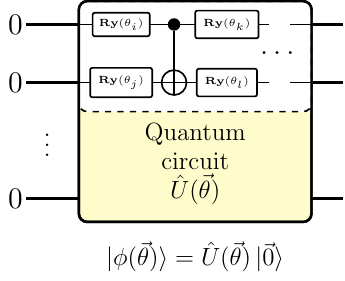}
	\caption{\label{fig:q-circ}Sketch of the quantum circuit ansatz 
		$\vert\phi(\vec{\theta})\rangle = \hat U(\vec{\theta})\vert\vec{0}\rangle$ 
		depending on a set of parameters, $\vec{\theta}$, in form of single qubit rotations around the $y$-axis, $\mathbf{\rm Ry(\theta_i)}$. 
	}
\end{figure}
An alternative to unitary approximation is to express QITE in variational form -- VarQITE\cite{mcardle_variational_2019}. 
In VarQITE, one approximates the targeted state $\ket{\psi}$ with a quantum circuit ansatz, $\hat U(\Vec{\theta})$, that depends on a set of parameters $\Vec{\theta}$ with elements $\theta_i$, i.e. representing the angles of single qubit rotational gates (Fig.~\ref{fig:q-circ}),
\begin{equation} \label{eq:varqiteans}
	\ket{\psi(\tau)} \approx \hat U(\Vec{\theta}(\tau)) \ket 0 = \ket{\phi(\Vec{\theta}(\tau))} = \ket{\phi(\tau)}.
\end{equation}
The ITE can then be approximated using McLachlan’s variational principle \cite{mclachlan_variational_1964, Yuan2019}. 
This approach minimizes the distance between the QITE evolution and the approximated path in parameter space $\hat U(\vec{\theta})$,
\begin{equation} \label{eq:mclachvar}
	\delta \left| \left(\pdv{}{\tau} + \Hat{H} - E_\tau\right) \ket{\psi(\tau)} \right| = 0,
\end{equation}
where $\big|\ket{\psi}\big| = \sqrt{\bra{\psi} \ket{\psi}}$. 
Equation \eqref{eq:mclachvar} minimizes the distance between the left-hand- and right-hand-side of the Wick-rotated Schrödinger equation, Eq.~\eqref{eq:wick}, 
and the energy expectation value at imaginary time $\tau$.
$
E_\tau = \bra{\psi(\tau)} \Hat{H} \ket{\psi(\tau)},
$
ensures normalization.\cite{mcardle_quantum_2020}
VarQITE iteratively steps through imaginary time, approximating the ideal path of QITE. 
The update rule for each iteration can be obtained from first expanding Eq.~\eqref{eq:mclachvar} in the parameter space by inserting Eq.~\eqref{eq:varqiteans}, which simplifies to~\cite{mcardle_variational_2019, Yuan2019}
\begin{equation} \label{eq:updaterule}
	\sum_j A_{ij} \Dot{\theta}_j = C_i.
\end{equation}
In Eq.~\eqref{eq:updaterule}, $\Dot{\Vec{\theta}}$ with elements $\Dot{\theta}_j$ represents the imaginary time derivative of the quantum circuit parameters $\Vec{\theta}$ (Fig.~\ref{fig:q-circ}), and
\begin{equation} \label{eq:AC}
	\begin{split}
		A_{ij} =& \Re \left( \pdv{\bra{\phi(\tau)}}{\theta_i} \pdv{\ket{\phi(\tau)}}{\theta_j} \right),\\
		C_i =& \Re \left( - \pdv{\bra{\phi(\tau)}}{\theta_i} \Hat{H} \ket{\phi(\tau)} \right).
	\end{split}
\end{equation}
The matrix with elements $A_{ij}$ and vector with elements $C_i$, which both depend on the imaginary time $\tau$, are the metric tensor in parameter space $\mathbf{A}(\tau)$~\cite{Fubini1908, Study1905} related to the quantum Fisher information matrix \cite{stokes_quantum_2020, Gacon2021,  Yao2022, Wilczek1989, Hackl2020, zhoujiang2019, Liu2020QuantumFisherinformation, Giovannetti2011, Petz1996} and the gradient $\Vec{C}(\tau)$, respectively.
From Eq. \eqref{eq:updaterule}, one can solve for $\Dot{\Vec{\theta}}$ by inverting $\mathbf{A}$, and then update $\Vec{\theta}$ by i.e. the Euler or Runge-Kutta methods\cite{Zoufal2023}.

\begin{figure}
	\includegraphics[width=0.4\textwidth]{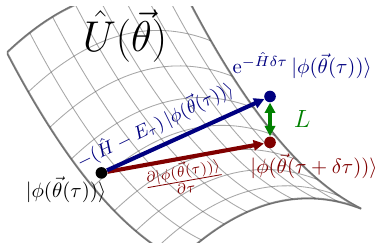}
	\caption{\label{fig:sketch}VarQITE is an approximation to QITE. The McLachlan distance, $L$ (green) quantifies how far a point on the manifold given by the ansatz $\hat U(\Vec{\theta})$ (red) is from the ideal QITE path (blue). VarQITE minimizes this distance after each small imaginary time step $\delta \tau$.
	}
\end{figure}

The drawback of the VarQITE approximation compared to QITE is that the strict convergence guarantee to the ground state is lost, as one is limited by how expressive the employed ansatz $\hat U(\vec{\theta})$ is.
When quantifying how close the iterative VarQITE is to the QITE path, it is helpful to consider the McLachlan distance $L$, \cite{Yuan2019}
\begin{equation} \label{eq:McLachDist}
	L = \sqrt{\sum_{i,j} A_{ij} \Dot{\theta}_i \Dot{\theta}_j - 2 \sum_i C_i \Dot{\theta}_i + 2\, \text{Var}(\Hat{H})}.
\end{equation}
Figure~\ref{fig:sketch} illustrates how the quantity $L$ can be interpreted as the distance between the optimal path of QITE and the approximate path of VarQITE~\cite{Yuan2019}. We will return to describe why $L$ is particularly important for TC-AVQITE when introducing our method of choice for adaptive ansatz construction.

The cost in terms of circuit evaluations for measuring $\mathbf{A}$ on quantum hardware scales as $\mathcal{O}(n_{\theta}^2)$, where $n_{\theta}$ is the number of ansatz parameters.
Fortunately, various approximations are available that reduce this scaling to linear,~\cite{stokes_quantum_2020, Fitzek2023, Gacon2023-jg, Gacon2023-hh} or even a constant cost.~\cite{Gacon2021}
However, it was recently shown by van Straaten and Koczor~\cite{Straaten2021} that the measurement cost of the gradient will dominate for large-scale quantum chemistry applications.

\subsection{Adding adaptive ansätze}

Adaptive ansätze are iteratively built to identify a circuit that is as shallow as possible yet sufficiently deep to describe a given problem. Adaptive quantum algorithms gradually add operators from a pre-defined operator pool to an initial, easy-to-prepare ansatz circuit. Operator pools can be constructed in various ways.\cite{Dyke2024, Shkolnikov_Mayhall_Economou_Barnes_2023} For example, they might include fermionic excitation operators \cite{grimsley_adaptive_2019} or operators constructed from Pauli strings \cite{tang_qubit-adapt-vqe_2021}. Which operator(s) to append and when to do so is decided iteratively based on some selection and expansion criteria. Using an adaptive approach naturally decreases circuit depth compared to a case in which all operators in the pool are used.

The first adaptive ansatz implementation was Adaptive Derivative-Assembled Problem-Tailored
ansatz (ADAPT)-VQE by Grimsley \textit{et al.} \cite{grimsley_adaptive_2019}, which has been followed by several variants.~\cite{tang_qubit-adapt-vqe_2021, gomes_adaptive_2021, Grimsley2023, Feniou2023, Burton2023} Herein, we rely on the adaptive algorithm AVQITE by Gomes \textit{et al.} \cite{gomes_adaptive_2021}, which implements adaptive algorithms in the context of VarQITE.

In AVQITE as implemented in [\onlinecite{gomes_adaptive_2021}] and in our work, the operator pool consists of all Pauli strings of a unitary coupled cluster singles doubles (UCCSD) ansatz\cite{Anand2022} constructed for the problem. The AVQITE ansatz circuit is expanded by selecting those operators that keep the VarQITE evolution as close to QITE as possible. Operators are added when the McLachlan distance $L$, Eq.~\eqref{eq:McLachDist}, becomes too large compared to some defined cutoff value $L_\text{cut}$. 

Adaptive ansätze have been shown to successfully decrease the circuit depth compared to including the entire operator pool at the cost of more measurements. The reason for these extra measurements is that the adaptive algorithm needs to evaluate the expansion and selection criteria to keep track of when to modify the ansatz circuit. However, as circuit depth is currently one of the most limiting factors for NISQ hardware, there is much to be gained from the approach despite the increased measurement cost. Additionally, work has been done to reduce measurement costs~\cite{Shkolnikov_Mayhall_Economou_Barnes_2023, Anastasiou2023-vj, Ramoa2024-oe}, for example, through classical shadows~\cite{Huang_Kueng_Preskill_2020}, Pauli grouping~\cite{Yen_Ganeshram_Izmaylov_2023, Yen2020, Yen2021, Izmaylov2019}, and informationally complete positive operator valued measures~\cite{Lan2022, Sapova2022, Liu2021, nykanen2023mitigating, Fischer2024-kt}.

\subsection{The transcorrelated method}

\begin{figure*}
	\includegraphics[width=\textwidth]{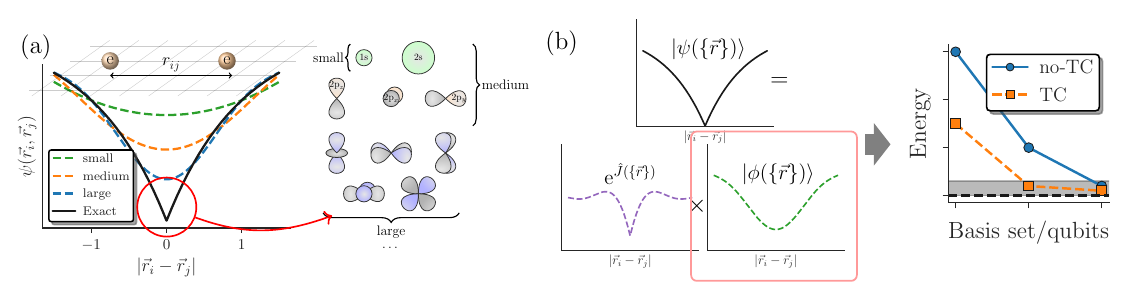}
	\caption{\label{fig:tc-motivation}(a) Sketch of how the cusp of the electronic wavefunction, $\psi(\{\Vec{r}\})$, necessitates the use of large Gaussian-type orbital basis sets. 
		(b) Sketch of how factoring the electronic wavefunction in Jastrow form, $\ket{\psi(\{\Vec{r}\})} = \textrm{e}^{\hat J}  \ket{\phi(\{\Vec{r}\})}$, where $\textrm{e}^{\hat J}$ captures the cusp, leads to better results for $\ket{\phi(\{\Vec{r}\})}$ in smaller basis sets.}
\end{figure*}

The TC method introduced by Hirschelder~\cite{Hirschfelder1963}, Boys and Handy~\cite{Boys1969, Handy1969, Handy1969b, BoysHandy1969}, is an explicitly correlated method~\cite{Hylleraas1929, Httig2011, Kong2011, Tenno2011, Tenno2012, Grneis2017, Kutzelnigg1985, Tenno2004, Tenno2004b} based on factorizing the electronic wave function in Jastrow form \cite{jastrow_many-body_1955},  
\begin{equation}
	\label{eq:j}
	\ket{\psi} = e^{\hat J} \ket{\phi}, \quad \hat J = \sum_{i<j} J_{ij} u(\Vec{r_i}, \Vec{r_j}),
\end{equation}
where $u$ is a symmetric correlation function over electron pairs and $J_{ij}$ are optimizable parameters. 
Eq. \eqref{eq:tc_transformation} allows us to recast the stationary Schrödinger equation, Eq.~\eqref{eq:schrodinger-stat}, in terms of $\ket{\phi}$,
\begin{align}
	% \begin{split}
		\Hat{H} \ket{\psi} = E \ket{\psi} \quad \Rightarrow  \quad &\Hat{H} e^{\hat J} \ket{\phi} = E e^{\hat J} \ket{\phi} \\ %\Rightarrow\\
		\Rightarrow \quad \underbrace{e^{-\hat J} \Hat{H} e^{\hat J}}_{\Bar{H}} \ket{\phi} &= E \ket{\phi}  % \Rightarrow \Bar{H} \ket{\phi} = E \ket{\phi}.
		%  \end{split}
	\label{eq:tc_transformation}
\end{align}
Note that Eq.~\eqref{eq:tc_transformation} is not an approximation but an exact similarity transformation of the electronic Hamiltonian, 
Eq.~\eqref{eq:2bodyterms}\cite{dobrautz_compact_2019}. 
However, and importantly, this recasting is \textit{not} a unitary transformation, so Hermiticity is lost\cite{Cohen2019}. The \enquote{normal} Rayleigh-Ritz variational principle requires Hermiticity, which means many variational methods, such as the VQE, do not apply to (non-truncated) transcorrelated Hamiltonians. 
There are ways to perform an approximate Hermitian truncation of a transcorrelated similarity transform if Hermiticity is deemed essential, 
	i.e. via CT-F12 theory\cite{Neuscamman2010, Yanai2012} used by Motta \textit{et al.}~\cite{Motta2020} and Kumar \textit{et al.}\cite{Kumar2022}
However, as a projective method, VarQITE can be used with both non-Hermitian as well as Hermitian Hamiltonians\cite{Mcardle2020, sokolov_orders_2022}.

In quantum chemistry, explicitly correlated methods are essential for correctly dealing with Kato's cusp condition: \cite{Kato1957, pack_cusp_1966} that when two electrons approach each other, they (should) give rise to a sharp, non-differentiable dip (cusp) in the wave function (Fig.~\ref{fig:tc-motivation}a). 
This sharp feature is one of the reasons why large basis sets are needed in conventional quantum chemistry calculations.
As basis sets are generally composed of smooth functions, such as Gaussians, many basis functions are needed to capture this cusp (Fig.~\ref{fig:tc-motivation}a).  
However, through the TC method, the cusp condition can be directly treated by choosing an appropriate Jastrow factor $\Hat{J}$ (see Eq.~\ref{eq:j} and Fig.~\ref{fig:tc-motivation}b), so that the non-differentiable behaviour of $\ket{\psi}$ can be incorporated in the Hamiltonian $\Bar{H}$. The cusp condition description has then moved from the wave function into the Hamiltonian via the similarity transformation, as seen in  Eq.~\eqref{eq:tc_transformation}.

Dealing with the cusp condition in the Hamiltonian instead of the wavefunction is why the TC method can provide results much closer to the complete basis set (CBS) limit with smaller basis sets (Fig.~\ref{fig:tc-motivation}b). This potential for lowering computational resources is one reason behind the TC methods' recent revival in electronic structure theory.~\cite{dobrautz_compact_2019, Cohen2019, Guther2021, Baiardi2020, Baiardi2022, Dobrautz2022, Jeszenszki2020, Liao2023, Liao2021, Schraivogel2023, Schraivogel2021, Ammar2022, Giner2021, Haupt2023, Tenno2023, Ochi2023, Ochi2023a, Ammar2023-lo, Lee2023-oo} 
Additionally, a smaller basis set means that one needs significantly fewer qubits to obtain reliable and accurate quantum chemistry results, 
as the number of qubits scales with active space size.~\cite{Motta2020, schleich2021improving, Mcardle2020, Kumar2022, sokolov_orders_2022, dobrautz_ab_2023} 
This resource reduction has recently been demonstrated by Motta \textit{et al.}\cite{Motta2020} and Kumar and coworkers\cite{Kumar2022} using CT-F12 theory, Schleich \textit{et al.} using the $[2]_{\rm R12}$ correction and by some of the present authors with an un-approximated TC-VarQITE combination\cite{sokolov_orders_2022, dobrautz_ab_2023}.

Moving the description of the cusp from the wave function to the Hamiltonian does not come for free. The price we pay is that the Hamiltonian becomes more complex, both in terms of the aforementioned non-Hermiticity (in the form of modified two-body terms) as well as the appearance of three-body terms, see Ref.~[\onlinecite{Cohen2019}] for details. 
%-- compare Eqs.~\eqref{eq:2bodyterms} and \eqref{eq:3bodyterms}:

%\begin{equation}
%    \begin{split}
	%        \Bar{H} =& \underbrace{\sum_{pq} h_{p}^q a^\dagger_p a_q}_\text{one-body terms} + \underbrace{\frac{1}{2} \sum_{pqrs} (V_{pq}^{rs}-K_{pq}^{rs}) a^\dagger_p a^\dagger_q a_r a_s}_\text{two-body terms} -\\
	%        &- \underbrace{\frac{1}{6} \sum_{pqrstu} L_{pqr}^{stu} a^\dagger_p a^\dagger_q a^\dagger_r a_s a_t a_u}_\text{three-body terms}.
	%    \end{split}
%    \label{eq:3bodyterms}
%\end{equation}
%In Eq. \eqref{eq:3bodyterms}, $K$ and $L$ are WHAT-To-Call-them that arise from the TC similarity transformation, and are dependent on the choice of $u(\Vec{r_i}, \Vec{r_j})$\ as introduced in Eq.~\ref{eq:j}. 

The additional three-body terms of the TC method raise the justified question of whether applying the method will be beneficial at all -- does the incurred cost outweigh the benefits? Recent work~\cite{Dobrautz2022, Haupt2023, Christlmaier2023} has shown 
that the TC Hamiltonian can be reduced to an $\bigO{N^5}$ 
or even $\bigO{N^4}$ scaling (with $N$ being the number of orbitals) by either neglecting three-body excitations with six unique indices\cite{Dobrautz2022}) altogether or by neglecting the pure normal ordered\cite{Kutzelnigg1997} three-body operators and incorporating the remaining three-body contributions in the two-, one-, and zero-body integrals.~\cite{Christlmaier2023} 
Additionally, in Ref.~[\onlinecite{dobrautz_ab_2023}], some of the present authors have demonstrated that the resource reduction of the TC method (without approximations) outweighs the cost of additional measurements until (roughly) the 1000 qubit mark. 

\subsection{\label{sec:number}Conserved quantum numbers}

The Jastrow factor $\e^{\hat J}$, Eq.~\eqref{eq:j},  used in the TC approach is optimized for a state with a specific chosen number of electrons, $n_{\rm mol}$,\cite{Haupt2023} usually corresponding to the molecular ground state. 
Under certain extremal conditions, e.g., in the broken bond regime, the TC similarity transformation, Eq.~\eqref{eq:tc_transformation}, can cause sectors of the Hamiltonian describing different electron numbers, which should have higher energy, to be below the original ground state sector with $n_{\rm mol}$ electrons.
%In a Fock space picture of quantum chemistry, states with higher electron numbers can thus be below the true ground state. 
Thus, unless measures are taken to conserve the correct number of electrons, TC calculations may converge to false ground states.

These Hamiltonian symmetry sector issues can be avoided by modifying the energy calculation to include a penalty term, $E' = E + E_{\rm penalty}$.
The energy penalty is in this work given by \begin{equation}\label{eq:penalty}
	E_\text{penalty} = \alpha \bra{\phi(\vec\theta)} (\Hat{N} - n_{\rm mol})^2 \ket{\phi(\vec\theta)},
\end{equation}
where $\Hat{N}$ is the electron number operator, $n_{\rm mol}$ is the chosen number of electrons, and $\alpha$ is a constant we have set to 1.
Equation~\eqref{eq:penalty} penalizes solutions with an electron number different than the chosen, $n \neq n_{\rm mol}$, by increasing their respective energy expectation value $E'$.
This penalty term ensures that TC-AVQITE converges to physically/chemically sound solutions with the correct number of electrons, $n_{\rm mol}$, but does not affect the described physics or chemistry of the studied systems.

\section{\label{sec:method}Computational details}

\begin{figure}
	\centering
	\includegraphics[width=0.7\linewidth]{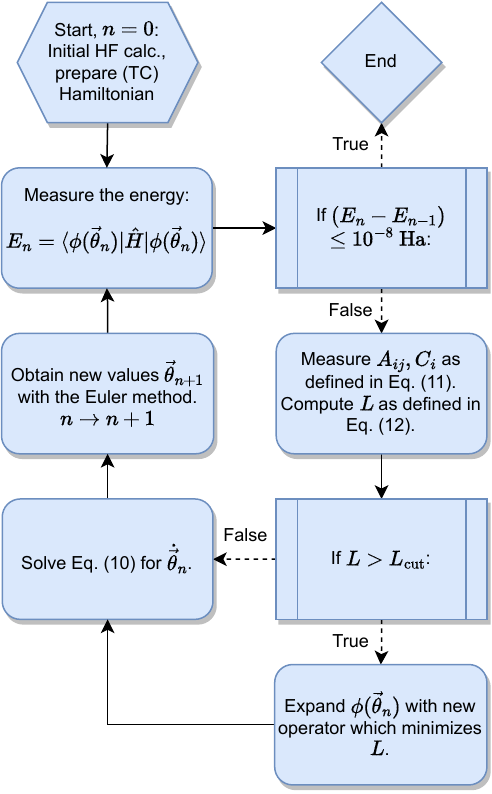}
	\caption{The iteration procedure for TC-AVQITE.}
	\label{fig:flowchart}
\end{figure}

The combination of the TC method with AVQITE follows naturally by applying each method sequentially: the TC method first produces a Hamiltonian $\Bar{H}$, which can then be used in a modified AVQITE implementation. 
Our program for performing TC-AVQITE is based on the code by Gomes \textit{et al.}'s code, as implemented in Refs.~\onlinecite{gomes_adaptive_2021, yao_avqite_2022}. 
Our development, available as a Python code,\cite{dobrautz_avqite_2023} includes additions to handle non-Hermiticity, generation of Hamiltonians and appropriate operator pools and restartable calculations. 
Our implementation of TC-AVQITE relies on Qiskit v.0.42.0, Qiskit Nature v.0.5.2, \cite{qiskit_contributors_qiskit_2023, developers_qiskit_2023} and Qutip v.4.7.1 \cite{j_r_johansson_qutip_2013} to produce operator pools and obtain exact reference energies. 
TC Hamiltonians were generated following the workflow outlined in Refs.~\onlinecite{dobrautz_ab_2023, Haupt2023}. 
PySCF v.2.4.0 \cite{sun_recent_2020, Sun_2018, Sun_2015} was used for the initial Hartree-Fock calculations to construct the molecular orbital basis for generating the Hamiltonians and operator pools, as well as the FCI/cc-p(C)VTZ and FCI/CBS estimate calculations. 

We have performed noiseless, state-vector AVQITE and TC-AVQITE  simulations to compute the ground state energy of the three molecular test cases. 
To limit costs associated with computing the six-body integrals in the TC approach and the state-vector simulations, we have used a minimal STO-6G basis set in all simulated systems. 
We note that more elaborate basis sets are required to approach the CBS, even in TC-based approaches. 
As such, a cost-effective alternative would be using an active space approach combined with a larger basis set. 
The use of minimal basis sets suffices for our goals here: to demonstrate the circuit width and depth reduction made possible with TC-AVQITE.

In our calculations of the water molecule, the oxygen 1s orbital was frozen, i.e., omitted from the correlated description, which resulted in a reduction of two qubits.
To decrease calculation costs further, parity encoding \cite{seeley_bravyi-kitaev_2012} was used for all systems to decrease the number of simulated qubits by two. 

The TC-AVQITE iteration procedure is visualized in Fig. \ref{fig:flowchart}:
First, we perform a conventional HF calculation, and in the case of TC-AVQITE, we use the workflow of Ref.~[\onlinecite{Haupt2023}] to construct the TC Hamiltonian. 
In all cases except for broken-bond H$_2$O, we use the single determinant HF state as the initial state, $\vert \phi(\vec{\theta}_0)\rangle$.  
Since the HF solution is not a good approximation to the true ground state for H$_2$O at \SI{2.5}{Å}, we use a single determinant open-shell initial state in this case. 
%The initial states were chosen as the states with the largest overlap with the conventionally calculated ground state. 
%Except in the broken-bond regime of H$_2$O, this corresponded to the Hartree-Fock ground state. 
Then, we enter the TC-AVQITE self-consistent loop by measuring the energy expectation value of the current ansatz circuit.
%with an initially trivial ansatz circuit, and measure its energy expectation value. 
If this expectation value does not change by more than \SI{1e-8}{Ha} during the iteration procedure, 
we consider the calculation converged and exit the loop.
If not, we measure the metric tensor with elements $A_{ij}$ and the gradient with elements $C_i$, and from these, compute the McLachlan distance $L$. 
If the McLachlan distance is larger than the defined cutoff value, $L > L_\text{cut}$, we expand the ansatz circuit with a new operator. 
We use the operator with the largest decrease to $L$ when added. 
This step is skipped if $L \leq L_\text{cut}$. 
Next, we solve Eq.~\eqref{eq:updaterule} to obtain $\Dot{\Vec{\theta}}$, and obtain new parameter values $\Vec{\theta}$ by the Euler method. The loop is repeated until the energy convergence criterion is met. 

%For details on initial state selection, see the supplementary information (SI).~\cite{SI}

We used $\Delta \tau = 0.05$ as the imaginary time step and $L_\text{cut} = \SI{1e-5}{}$ for the McLachlan distance cutoff in all AVQITE and TC-AVQITE calculations. 
Both TC-AVQITE and the original AVQITE algorithm by Gomes \textit{et al.}\cite{gomes_adaptive_2021} were tested and found to be robust with respect to different parameter settings, details of which can be found in the Supporting Information (SI)\cite{SI}. 
TC-VarQITE calculations were performed as outlined in Ref.~\onlinecite{dobrautz_ab_2023} using the same imaginary time step of $\Delta \tau = 0.05$. 

\textbf{Notes on convention:} We follow the convention of Ref. \onlinecite{lolur_reference-state_2023} and use the term \textquote{computational accuracy} in this work when the difference of a quantum calculation and the exact solution in a given (finite) basis set do not exceed \SI{e-3}{Ha}. We highlight this nomenclature because a different term, \textit{chemical} accuracy, is often used in quantum chemistry to measure calculation quality. Chemical accuracy is commonly defined as an error of \SI{1}{kcal/mol} (approximately \SI{e-3}{Ha}) with respect to the exact (e.g., FCI/CBS) solution or experiment \cite{lolur_reference-state_2023}. 
In quantum computing literature, this term is sometimes used instead of what we prefer to call computational accuracy. Such mix-up can be misleading -- especially as these energies usually differ significantly \cite{comparison_nist_2022}. 

Furthermore, to avoid confusion, we want to make clear how we use the terms \textit{FCI} and \textit{ED} in the remainder of the text:
Both terms, FCI and ED, are somewhat interchangeably used in computational chemistry and physics. 
They refer to a given Hamiltonian's exact ground state (energy) solution expressed in a specific basis set. 
However, as stated above, the main benefit of explicitly correlated/TC approaches is that they usually yield lower energies in a given basis set. 
To avoid confusion, we refer to the energy obtained by exactly diagonalizing the TC Hamiltonian (Eq.~\eqref{eq:tc_transformation}) as the ED -- TC result (here only performed in the STO-6G basis).
Meanwhile, we resort to the \enquote{usual} convention FCI/basis when referring to the exact solution of the original Hamiltonian (Eq.~\eqref{eq:2bodyterms}) in a specific basis set. 
We want to note that these two energies agree in the CBS limit.  

\begin{figure*}
	\centering
	\includegraphics[width=\linewidth]{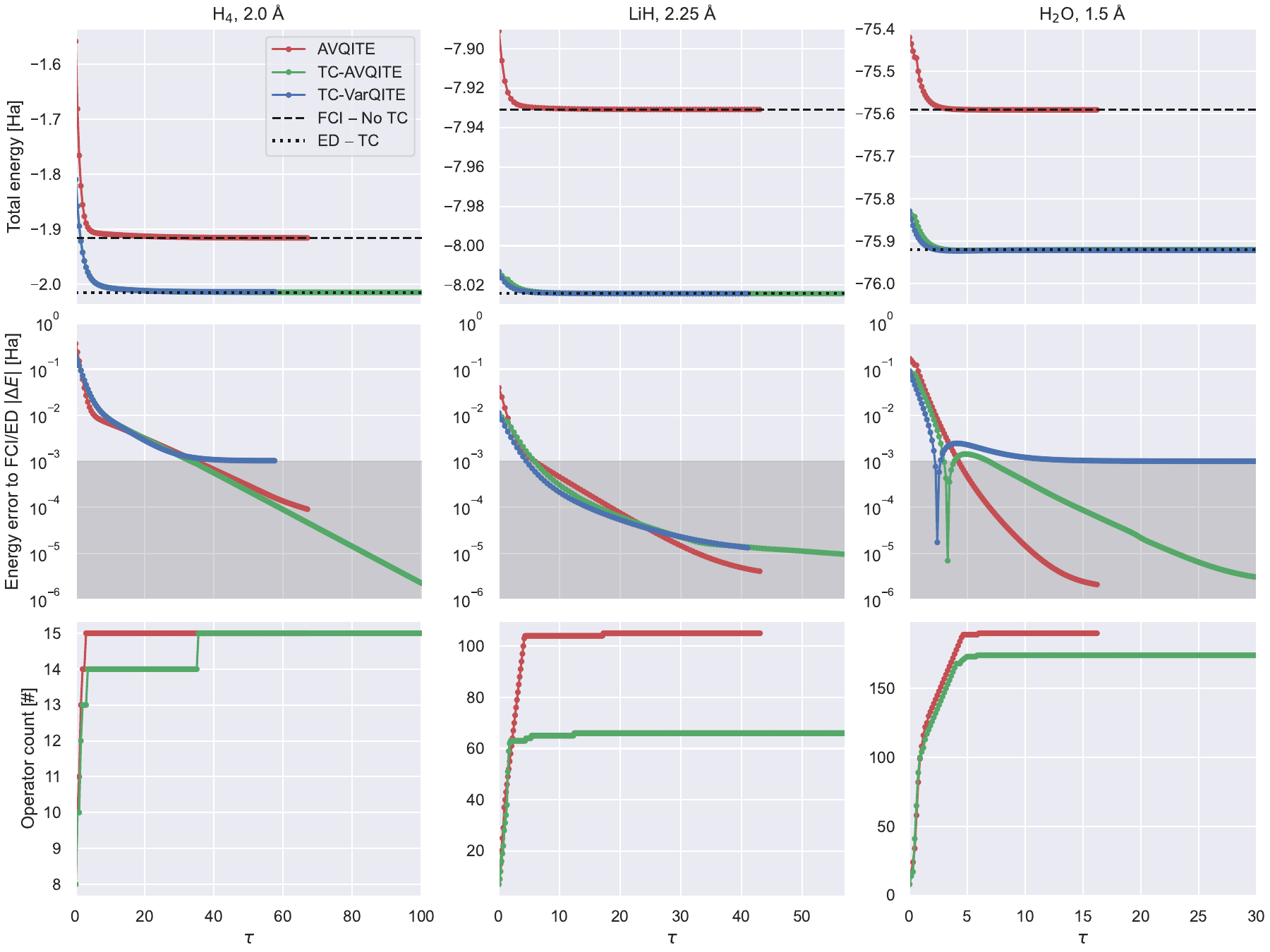}
	\caption{TC-AVQITE (green), AVQITE (red) and TC-VarQITE (blue) evolutions for H$_4$ (left), LiH (middle) and H$_2$O (right column) in the half-broken bond regime using a STO-6G basis set. 
		Top row: Total energy vs. imaginary time $\tau$. 
		Middle row: Energy error of AVQITE relative to FCI/STO-6G and TC-AVQITE/TC-VarQITE relative to the ED of the TC Hamiltonian in the STO-6G basis. 
		The shaded area indicates computational accuracy. 
		Note the sharp discontinuity of the energy error for H$_2$O -- this feature arises when the TC-AVQITE/TC-VarQITE energy estimate crosses the ED -- TC result.
		Bottom row: Number of adaptively added operators vs. imaginary time.     
	For clarity of demonstration, the number of operators used by TC-VarQITE (the full UCCSD pool) is omitted. 
			These numbers are 152 for H$_4$ and 640 for LiH and H$_2$O, respectively.
	}
	\label{fig:3x3evolutions}
\end{figure*}

\section{Results and Discussion}\label{section:results}

%\subsection{Cutting circuit depth by cutting circuit width} \label{sec:tc-vs-notc}

One assumption motivating our development of TC-AVQITE is that by explicitly dealing with Kato's cusp condition with a TC transformation\cite{dobrautz_compact_2019, dobrautz_ab_2023}, total energies should reach closer to the CBS limit compared to AVQITE. 
Furthermore, we expect that transferring complexity from the wavefunction to the Hamiltonian in the TC method (Eq.~\eqref{eq:tc_transformation}), should translate to shallower quantum circuits.\cite{sokolov_orders_2022} 
In other words, the two metrics relevant for comparing TC-AVQITE and AVQITE are the computed energies and the number of operators adaptively appended to the ansatz circuits. 

To compare TC-AVQITE with AVQITE, we study bond dissociation in three test cases: quadratic H$_4$, LiH and H$_2$O. 
In the latter case, the bond dissociation is defined with respect to symmetric stretching of the $r_{\rm OH}$ distances at a fixed angle of $\angle(\text{HOH}) = 104.4\degree$. This test set was chosen so to include different kinds of chemical bonds, including ionic (LiH) and polar covalent (H$_2$O), as well as to stress-test the methodology in strongly correlated systems (H$_4$). 
In what follows, we study different points along these systems' potential energy surfaces to capture behaviours of bonded, broken bond, and half-broken bond regimes. For benchmarking purposes, all TC-AVQITE calculations are additionally compared to FCI computations performed with different basis sets (STO-6G, cc-p(C)VTZ or CBS-limit extrapolation) and our original TC-VarQITE implementation\cite{sokolov_orders_2022, dobrautz_ab_2023} 
	using a full UCCSD ansatz.

\begin{table*}\small 
	\caption{\label{tab:circuits} 
			Estimates of final quantum circuit requirements and final energy errors for the calculations shown in Fig.~\ref{fig:3x3evolutions}. 
			All results use parity encoding with a subsequent 2-qubit reduction. We report the (final) number of used UCC operators and the corresponding required total number of gates, the number of CNOTs (obtained with Qiskit’s \texttt{count\_ops()} function), and the circuit depth.
	}
	\begin{ruledtabular}
		\begin{tabular}{lcccccccc}
			System & \#Qubits & Method & \#Operators &
			\#Gates & \#CNOTs & Circuit depth &
			\makecell{$|\Delta E|$ to STO-6G/ \\FCI or ED}
			& $|\Delta E|$ to CBS/FCI\\
			\colrule
			\multirow{3}{*}{H$_4$, \SI{2.0}{Å} } &  \multirow{3}{*}{6} & AVQITE & 15 & 271 & 118 & 161 & \SI{9.078e-5}{} & \SI{1.324e-1}{} \\
			&  & TC-AVQITE  & 15 & 269 & 116 & 151 & \SI{2.167e-6}{} & \SI{3.349e-2}{} \\ 
			&  & TC-VarQITE & 152 & 2797 & 1250 & 1643 & \SI{1.026e-3}{} & \SI{3.452e-2}{} \\ \hline
			\multirow{3}{*}{LiH, \SI{2.25}{Å}} & \multirow{3}{*}{10} &AVQITE & 105 &2993 & 1246 & 1476 & \SI{4.056e-6}{} & \SI{1.141e-1}{} \\
			&  & TC-AVQITE & 66 & 1752 & 704 & 842 & \SI{9.487e-6}{} & \SI{2.077e-2}{} \\ 
			&  & TC-VarQITE & 640 & 14570 & 7080 &  8468 & \SI{1.308e-5}{} & \SI{2.077e-2}{} \\ \hline
			\multirow{3}{*}{H$_2$O, \SI{1.5}{Å}} & \multirow{3}{*}{10} & AVQITE & 190 & 5724 & 2450 & 3069 & \SI{2.082e-6}{} & \SI{5.898e-1}{} \\
			&  & TC-AVQITE & 174 & 4792 & 2054 & 2537 & \SI{3.041e-6}{} & \SI{2.603e-1}{} \\ 
			&  & TC-VarQITE & 640 & 14572 & 7080 & 8460 & \SI{1.002e-3}{} & \SI{2.593e-1}{} \\
		\end{tabular}
	\end{ruledtabular}
\end{table*}

\subsection{Convergence in imaginary time}

\begin{figure*}
	\centering
	\includegraphics[width=\linewidth]{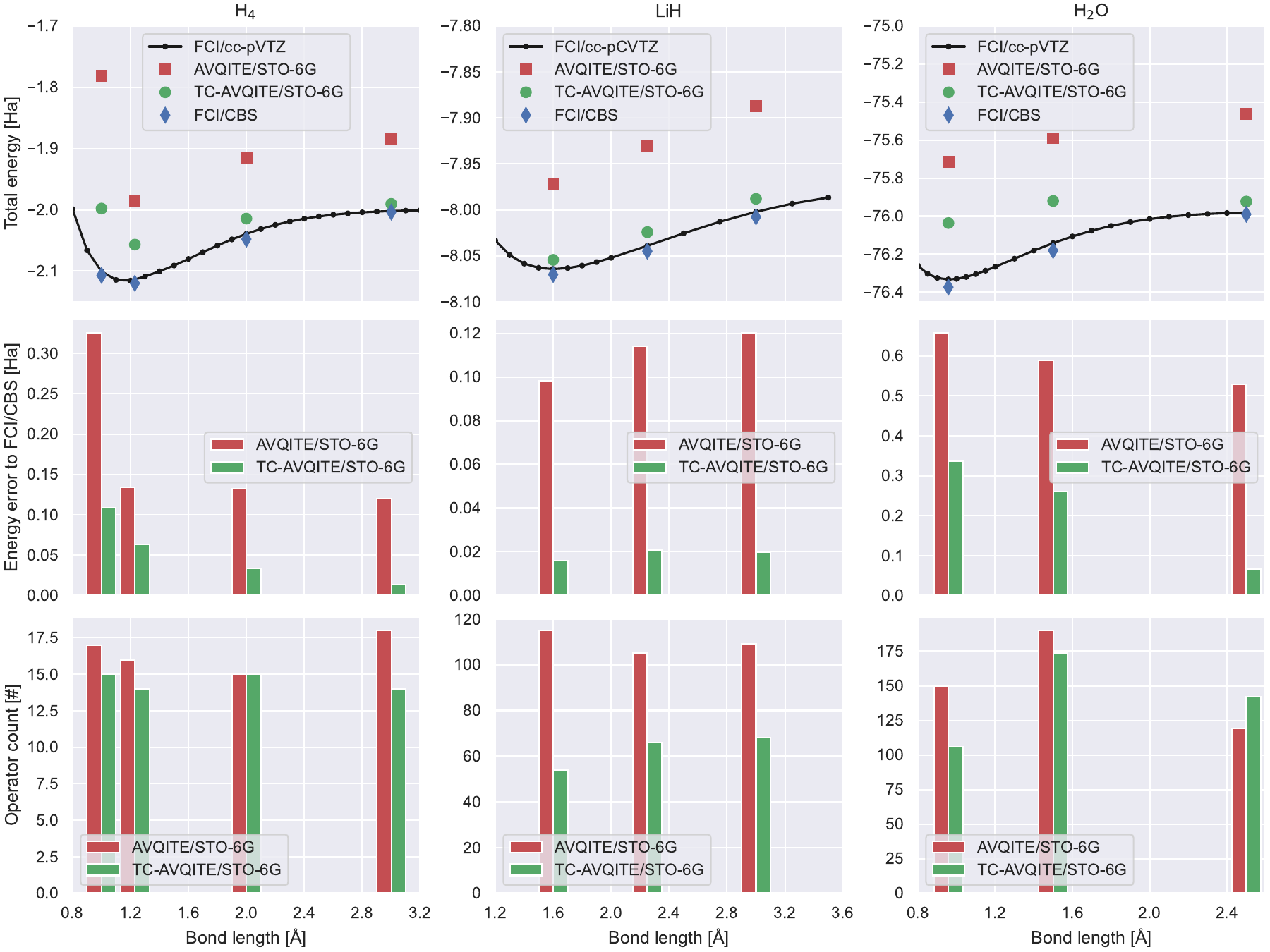}
	
	\caption{Comparison of results from TC-AVQITE/STO-6G (green), AVQITE/STO-6G (red), FCI/cc-p(C)VTZ (black) and FCI/CBS results (blue) for H$_4$ (left column), LiH (middle column) and H$_2$O (right column).
		(Top row) Total energy as a function of bond distance. (Middle row) Energy error with respect to FCI/CBS estimates as a function of bond distance. %; see SI\cite{SI} for details on CBS limit estimation.
		(Bottom row) The final number of adaptively added operators in AVQITE and TC-AVQITE calculations as a function of bond distance. Note: For visualization purposes, the bar plots in the middle and bottom rows are plotted as a function of bond distance. Adjacent pairs of red (AVQITE) and green (TC-AVQITE) bars correspond to the distances of the corresponding markers in the top row. Corresponding imaginary time trajectories are provided in SI~\cite{SI}.
	}
	%Note that the TC energies are lower than the CBS limit result for lithium hydride -- this can happen as the non-Hermiticity of the TC Hamiltonian means that the conventional Dirac-Frenkel variational theorem does not hold, i.e., the approximations are not guaranteed to be higher energy than the ground state.
	\label{fig:3x3properties}
\end{figure*}

Figure~\ref{fig:3x3evolutions} compares the imaginary time evolution of each tested molecule in the half-broken bond regime. 

The main strength of TC-AVQITE and TC-VarQITE is apparent from the first row of Fig.~\ref{fig:3x3evolutions}: There is a substantial difference between the conventional FCI result (FCI -- No-TC) and the energy obtained by exactly diagonalizing the TC Hamiltonian (ED -- TC), despite both using the same basis set. 
These improved energies due to the TC approach are the primary cause of the difference between TC-AVQITE/STO-6G and AVQITE/STO-6G when compared to FCI/CBS in Fig.~\ref{fig:3x3properties}, which we will discuss below.

The second and third rows of Fig.~\ref{fig:3x3evolutions} clearly illustrate how an initial rapid ramp-up of appended operators corresponds to a sharp decline in error. 
After this initial phase, the number of operators in the ansatz plateaus while the energy continues to converge steadily. 
One anomaly is apparent in the energy error for the calculation of H$_2$O. 
This sharp discontinuity arises because the energies of both TC-VarQITE and TC-AVQITE cross the reference energy obtained by exactly diagonalizing the TC Hamiltonian, expressed in the STO-6G basis set (ED -- TC). 
Such crossings can occur because the non-Hermiticity of the TC Hamiltonian $\Bar{H}$ invalidates the Rayleigh-Ritz variational theorem. 
In such situations, the TC-AVQITE and TC-VarQITE energy approaches the final ED -- TC value from below after the discontinuity. 
The final error of both adaptive methods (AVQITE and TC-AVQITE) is well below \textit{computational accuracy} concerning the corresponding reference values (FCI -- no-TC/ED -- TC).

	Opposed to TC-AVQITE, TC-VarQITE (using the entire operator pool) struggles to converge for H$_4$ at 2.0 \AA{} and H$_2$O at 1.5 \AA, where it retains an energy error to the ED -- TC result of around \SI{1}{mH}. 
	This behaviour is consistent with previous findings for ADAPT-VQE,\cite{grimsley_adaptive_2019} suggesting that adaptive ansätze can improve convergence.  
	In addition to the results shown here, 
	we compare TC-VarQITE with TC-AVQITE for all H$_4$ bond lengths, see the SI for details.\cite{SI}
	In general, our results indicate that TC-AVQITE is more accurate with respect to the ED -- TC energy than the TC-VarQITE calculation.
	The exception to this case is the broken-bond regime, where both perform roughly equivalently.

The third row of Fig.~\ref{fig:3x3evolutions} shows the lower operator count made possible with TC-AVQITE. 
For perspective and to appreciate the power of adaptive methods for reducing circuit depth without losing (and potentially even improving) accuracy, we note that the number of available operators (the full pool) that TC-VarQITE uses 
is 152 for H$_4$ and 640 for LiH and H$_2$O.

	Table.~\ref{tab:circuits} shows the final errors to FCI -- No-TC/ED -- TC results both to the STO-6G basis and to FCI/CBS results, as well as corresponding estimates of required quantum resources. 
	The listed resource estimates assume full circuit connectivity and are given as the total number of 1- and 2-qubit gates, the number of CNOTs, and the circuit depth (the number of gates that cannot be performed in parallel).
	Even for tiny target energy errors, $\abs{\Delta E} < 10^{-6}$, the required CNOT count for TC-AVQITE calculations is within reasonable limits of near-term quantum devices. 
	Due to the adaptive nature of TC-AVQITE, even lower CNOT counts are possible for less tight target $\abs{\Delta E}$. 
	In addition, specialised quantum circuited compilation/transpilation methods\cite{Hner2018, Earnest2021, Miller2024} could significantly reduce the required gate counts further.

	These results demonstrate that using adaptive quantum ansätze in TC-AVQITE  improves convergence and drastically reduces the number of CNOTs and circuit depth. 
	The circuit depth using TC-AVQITE is reduced by an order of magnitude for LiH and H$_4$ 
	and a factor of 5 for H$_2$O compared to TC-VarQITE.

\subsection{Bond dissociation}

Next, we compare TC-AVQITE with AVQITE along our test set's entire bond dissociation curves. 
We omit direct comparison with TC-VarQITE here, as both TC-AVQITE and TC-VarQITE target the same ED -- TC energies.
Figure ~\ref{fig:3x3properties} demonstrates a substantial advantage of TC-AVQITE, both in terms of lower total energies and fewer adaptively added operators.  
The error with respect to FCI/CBS results is up to an order of magnitude smaller using TC-AVQITE when applied to LiH across the entire binding curve, as well as in the stretched and broken-bond regimes of H$_4$ and H$_2$O. 
These drastically improved total energies are remarkable considering that only a minimal basis set is used, and clearly demonstrate the benefit of the TC method in reducing quantum circuit width (i.e., the number of necessary qubits). 

As shown in the bottom row of Fig.~\ref{fig:3x3properties}, this reduction in circuit width is accompanied by a simultaneous contraction of the required circuit depth. 
The benefit is modest for H$_4$ and H$_2$O, where the number of final operators is reduced by 8\% to 26\% on average, respectively. 
In the case of LiH, TC-AVQITE reduces the needed circuit depth by half compared to AVQITE while yielding results considerably closer to the CBS limit. 
However, there are exceptions. 
For example, for our calculation of quadratic H$_4$ with a side length of \SI{2.0}{Å}, both TC-AVQITE and AVQITE require 15 operators to reach convergence. 
We argue that the lack of improvement in situations such as these is due to a combination of (\textbf{a}) the minimal basis set size (only one spatial orbital per H atom); and (\textbf{b}) H$_4$ being a notoriously difficult system for unitary coupled cluster theory limited to single and double (UCCSD) excitations\cite{Sokolov2020}, which our operator pool is based upon. 

In the molecular disassociation limit of H$_4$ and H$_2$O, a curious behaviour can be observed: the TC-AVQITE energy errors to FCI/CBS are generally lower for larger bond lengths.  
This behaviour is most likely a basis set effect as the utilized minimal basis can better describe disassociated atoms than molecules. 
The presence of basis set effects motivates using larger basis sets, even with the TC method.

\begin{figure}
\centering
\includegraphics[width=\linewidth]{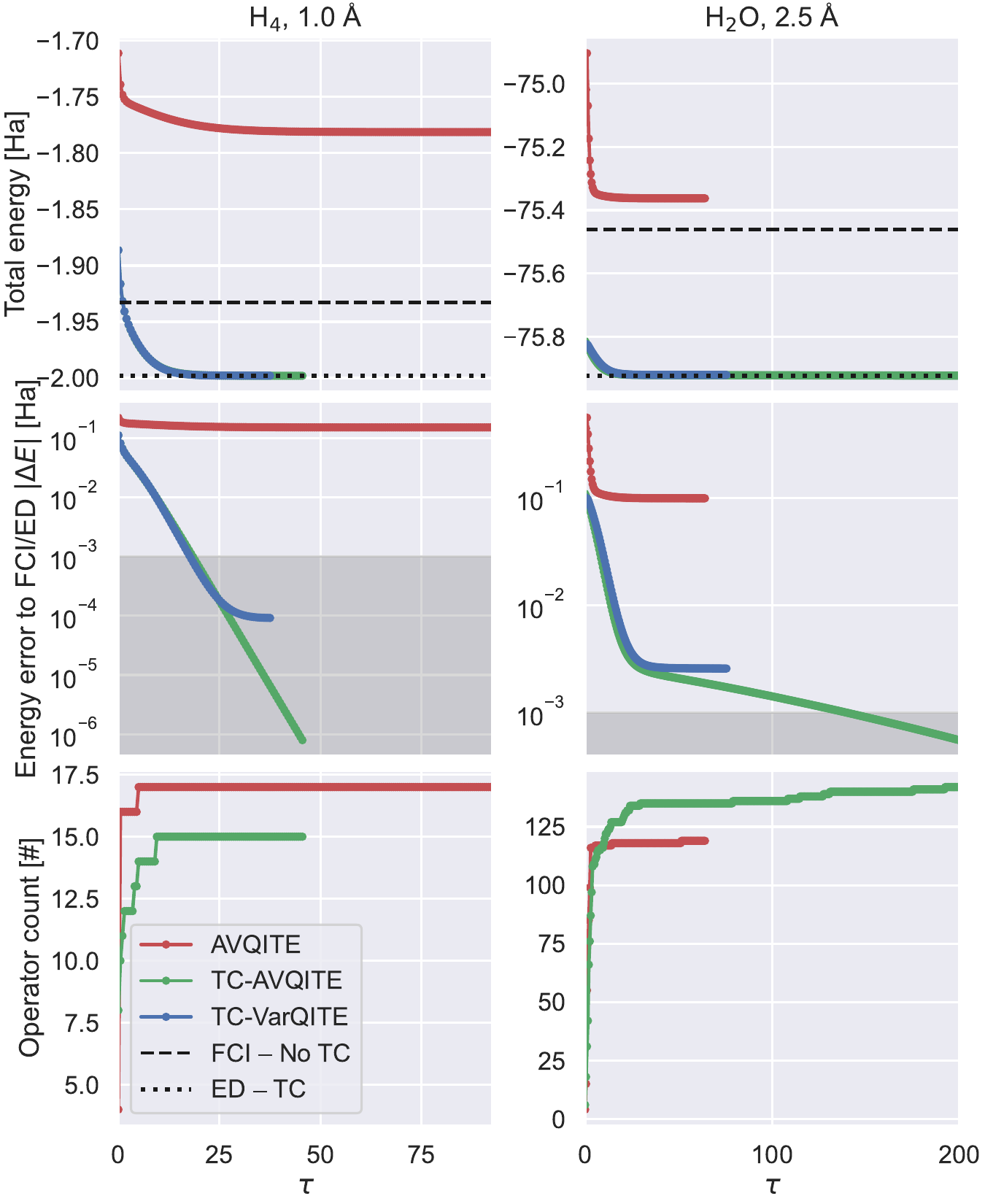}
\caption{Two examples where convergence was not reached for AVQITE: H$_4$ with bond length \SI{1.0}{Å} (left) and H$_2$O with bond length \SI{2.5}{Å} (right column), where TC-VarQITE also fails to converge, using a STO-6G basis. 
	Top row: A comparison between the TC-AVQITE, TC-VarQITE and AVQITE total energies. 
	Middle row: energy error of AVQITE relative to FCI/STO-6G and TC-AVQITE/TC-VarQITE with respect to ED of the TC Hamiltonian in the STO-6G basis. 
	Bottom row: operator counts for the two adaptive methods. }
\label{fig:special_case}
\end{figure}

In Fig.~\ref{fig:special_case}, we look more closely at two challenging examples where the operator count produced by TC-AVQITE might not look advantageous compared to AVQITE at first glance.  

First, in the case of the modest circuit depth reduction for quadratic H$_4$ at \SI{1.0}{Å}, TC-AVQITE adds something essential: in contrast to AVQITE, TC-AVQITE actually converges to the ground state solution! In contrast, AVQITE struggles to converge and retains a sizeable energy error exceeding \SI{0.1}{Ha} at convergence. The same convergence issue also occurs for AVQITE (though far less noticeably) when applied to H$_4$ at a bond length of \SI{3.0}{Å}, where computational accuracy cannot be reached; see the SI for details.~\cite{SI} By moving complexity from the wavefunction to the Hamiltonian, TC-AVQITE converges well below computational accuracy while, at the same time, requiring two operators less to do so. 

The second exception to circuit depth reduction seen in Fig.~\ref{fig:3x3properties} is H$_2$O at bond length \SI{2.5}{Å}.
In this case, the final operator count of TC-AVQITE is higher than the count for AVQITE, seemingly contradicting our assumption that TC approaches reduce circuit depth.  However, a closer look at the imaginary time evolutions (Fig.~\ref{fig:special_case} left column) reveals that similar to H$_4$ at \SI{1.0}{Å}, AVQITE here fails to converge to the FCI/STO-6G ground state. 
In other words, the high operator count for TC-AVQITE in these examples is caused by the algorithm successfully identifying suitable operators to append. If AVQITE were to converge in this example (for H$_2$O at \SI{2.5}{Å}), we expect its final operator count to be substantially larger. 

Similar as in the case for H$_4$ at 2.0 \AA{} and H$_2$O at 1.5 \AA{} (Fig.~\ref{fig:3x3evolutions}), TC-VarQITE also fails to converge for H$_2$O at 2.5 \AA{}. 
This case demonstrates that using adaptive ansätze in TC-AVQITE reduces circuit depth and improves convergence for strongly correlated systems like stretched H$_2$O.

The examples we have discussed highlight the strengths of the TC method -- how a similarity transformation can simplify the solution by transferring complexity from the wavefunction to the Hamiltonian. In other words, results from our test set of calculations support our premise that TC-AVQITE reduces both circuit width and depth, by yielding better results with smaller basis sets and shallower quantum circuits.
At the same time, TC-AVQITE can improve convergence behaviour compared to TC-VarQITE, which uses a full and pre-defined UCCSD operator pool.

\section{Conclusion} 

The most significant barrier to the practical quantum computation of chemistry is the performance of current hardware, which imparts particularly harsh restrictions on the quantum circuit width (number of qubits) and depth.
This work demonstrates that combining an explicitly correlated approach, the transcorrelated (TC) method, with adaptive quantum ansätze in the context of variational quantum imaginary time evolution (AVQITE) significantly reduces the necessary circuit depth and width. 

The TC method transfers complexity away from the wave function by incorporating Kato's cusp condition into the Hamiltonian description of a system. 
Consequently, the transformed TC Hamiltonian's eigenfunctions are easier to represent with smaller basis sets and require shallower adaptive quantum circuit ansätze. 
This quantum resource reduction enhances noise resilience and enables higher accuracy in calculating ground state energies for small molecular systems. 

By applying TC-AVQITE to the electronic structure problems of H$_4$, LiH, and H$_2$O, we demonstrate a close agreement with complete basis set limit results despite using a minimal basis set; in stark contrast to traditional (non-transcorrelated) methods. Additionally, we show that, by transferring complexity from the wavefunction into the Hamiltonian, TC-AVQITE is able to converge when applied to strongly correlated systems (H$_4$, stretched H$_2$O), where \enquote{conventional} AVQITE or non-adaptive approaches (TC-VarQITE) fail. 
Additionally, for the systems studied in this work, TC with adaptive quantum ansätze reduces the required quantum resources by a factor of 5 to 10 compared to a fixed quantum ansatz strategy employed in TC-VarQITE.

While the current study focuses on small molecular systems, we emphasise that TC-AVQITE holds promise for addressing larger, more complex quantum chemistry problems in the future. To achieve such up-scaling, we intend to combine the TC approach with more elaborate basis sets, active space approaches and self-consistent orbital optimization~\cite{Roos1980, Olsen2011, Dobrautz2021, fitzpatrick2022selfconsistent, deGraciaTrivio2023}, embedding methods,~\cite{Bauer2016, Tilly2021, Rossmannek2023, Rossmannek2021} as well as spin-conserving schemes.~\cite{Dobrautz2019b, Dobrautz2022b, LiManni2021, LiManni2020, Anselmetti2021, Burton2023} Because the TC method coupled with adaptive ansätze leads to more compact quantum circuits, the method should be inherently less susceptible to noise. Therefore, another future research direction will be to study the effect of hardware noise on the method's performance, moving closer to practical quantum chemistry application. 

\begin{acknowledgments}

We thank Dr. Yongxin Yao for sharing their AVQITE implementation, as well as providing helpful feedback. 

Funded by the European Union. 
Views and opinions expressed are, however, those of the author(s) only and do not necessarily reflect those of the European Union or European Research Executive Agency. 
Neither the European Union nor the granting authority can be held responsible for them.
This work was funded by the EU Flagship on Quantum Technology HORIZON-CL4-2022-QUANTUM-01-SGA project 101113946 OpenSuperQPlus100 and the Wallenberg Center for Quantum Technology (WACQT). WD acknowledges funding from the European Union’s Horizon Europe research and innovation programme under the Marie Sk{\l}odowska-Curie grant agreement No. 101062864. 
This research relied on computational resources provided by the National Academic Infrastructure for Supercomputing in Sweden (NAISS) at C3SE and NSC, partially funded by the Swedish Research Council through grant agreement no 2022-06725. We thank Mårten Skogh for valuable discussions and feedback.

\end{acknowledgments}

\section*{Additional Information}

Electronic Supplementary Information is available.~\cite{SI}
Data and software to reproduce this work will be available after peer review in an accompanying public Git repository.\cite{dobrautz_avqite_2023}

\end{document}